\documentclass[english,aps,floats,twocolumn,prb]{revtex4-1}
\makeatletter
\usepackage[latin9]{inputenc}
\usepackage{xcolor}
\usepackage{amssymb}
\usepackage{amsmath}
\usepackage{graphicx}

\makeatother

\begin{document}

\title{Impact of self-consistency in dual fermion calculations }

\author{T. Ribic$^{1}$, P. Gunacker$^{1}$ and K. Held$^{1}$
}

\affiliation{$^{1}$Institute of Solid State Physics, TU Wien, 1040 Vienna, Austria}

\date{\today}

\begin{abstract}
The dual fermion (DF) method allows for calculating corrections due to non-local correlations relative to an effective impurity  model. Choosing the impurity as that of a dynamical mean field theory (DMFT) solution at self-consistency is popular, and the corrections from dual fermion theory are physically meaningful. We investigate the effect of choosing the impurity instead in a self-consistent manner and find for the two dimensional Hubbard model an exponentially increase of the correlation length and susceptibility at low temperatures. There are pronounced differences for the two self-consistency schemes that are discussed in  the literature; the self-consistent DF solution can even be more metallic than the DMFT solution.
\end{abstract}


\maketitle
\let\n=\nu \let\o=\omega \let\s=\sigma \global\long\def\Gred{\mathcal{G}}
 \global\long\def\Chired{\overset{\sim}{\chi}}


\section{Introduction}

\label{Sec:Intro} Strongly correlated electron systems pose some
of the greatest challenges in modern solid state theory. The strong interplay between the electrons in such systems causes a multitude of interesting phenomena, such as superconductivity and interaction-driven metal-insulator transitions. While the underlying physics is interesting, the complexity also makes finding reliable or even numerical solutions notoriously hard.
Dynamical mean field theory (DMFT) \cite{Metzner1989,Georges1992a,Georges1996} has become a well established tool for treating purely local correlation effects. Based on this success,  multiple methods for extending DMFT and including non-local correlation effects as well have been  proposed. On the one hand there are cluster extensions of DMFT such as the dynamical cluster approximation (DCA) and cellular DMFT (CDMFT) \cite{Maier2005}.  On the other hand there are Feynman-diagrammatic extensions \cite{RMPVertex} such as the  dynamical vertex approximation (D$\Gamma$A) \cite{Toschi2007},  the dual fermion method (DF) \cite{Rubtsov2009}, the DMFT to functional renormalization group \cite{Taranto2014}, the non-local expansion scheme \cite{Li2015}, the one-particle irreducible approach (1PI) \cite{Rohringer2013}, and the  triply irreducible local expansion (TRILEX) \cite{Ayral2015}.

In this work, we study the two-dimensional Hubbard model on a square lattice with nearest neighbor hopping. At half-filling eminent questions are  antiferromagnetism, pseudogaps and the metal-insulator transition. From the Mermin-Wagner theorem \cite{Mermin1966}  we know that truly long-range antiferromagnetic ordering only sets in at zero temperature $T=0$, for perfect nesting there is antiferromagnetism  at arbitrarily weak interactions $U$. DMFT on the other hand gives a finite N\'eel temperature $T_N$ with mean-field critical behavior for the susceptibility and correlation length: $\chi \sim (T-T_N)^{-1}$, $\xi\sim (T-T_N)^{-0.5}$. 
One of the successes of the diagrammatic extensions of DMFT is to show instead   (around the $T_N$ of DMFT) a crossover to an exponentially increasing susceptibility\cite{Katanin2009,Otsuki2014}  $\chi \sim e^{a/T}$  and correlation length\cite{Schaefer2015-2}  $\xi \sim e^{b/T}$ with non-universal parameters $a$ and $b$.
This way the Mermin-Wagner theorem is eventually fulfilled with exponentially large correlation lengths instead of long-range order.

These long correlation length have, on the other hand, a strong impact on the metal-insulator transition. Here DMFT yields a first-order Mott-Hubbard metal-insulator transition with a second-order critical end-point \cite{Georges1996}, independent of dimension and hence also in two-dimensions.  Cluster DMFT with a finite $4\times 4$ momentum ($\mathbf k$) grid  also gives a  first-order transition, but with at a reduced $U_c$ and  opposite slope of the transition line because now the metallic instead of the insulating phase in DMFT has the larger entropy.\cite{Park2008} Similar results have been obtained using other methods that include  short range correlations only, such as the variational cluster approximation (VCA)~\cite{Schaefer2015-2} or second-order DF (DF$^{(2)}$).\cite{Loon2018b}

But in D$\Gamma$A \cite{Schaefer2015-2}, taking into account  long-range correlations of hundreds of sites,  the paramagnetic phase is always insulating for low enough temperatures, i.e., $U_c=0$.  The reason for this are the aforementioned long-range antiferromagnetic correlations. Even though there is no true antiferromagnetic order yet, the  exponentially large correlation length leads to a quasi-order so that the paramagnetic spectral function  has essentially the same gap as the antiferromagnetic phase. Due to perfect nesting on a square lattice Hubbard model, antiferromagnetism and hence the gapped low-$T$ paramagnetic state exist all the way down to $U_c=0$. A similar behavior is also observed in ladder DF \cite{vanLoon2017},  the two-particle self-consistent  theory (TPSC) \cite{Vilk1996,Vilk1997}, and the nonlinear sigma model approach \cite{Borejsza2003,Borejsza2004}.
 Also numerical results point in the same direction: extrapolated lattice quantum Monte Carlo (QMC) data\cite{Schaefer2015-2} show a similar insulating self-energy as  D$\Gamma$A; CDMFT \cite{Fratino2017} and DCA  \cite{Moukouri2001,WernerPC13,Merino2014} suggest at least a reduction of $U_c$ with increasing cluster size.
Against this trend, TRILEX yields an even larger $U_c$  than DMFT \cite{Ayral2016}.

Let us emphasize that the physics  of the metal-insulator caused by long-range antiferromagnetic spin fluctuations is distinctively different from that of Mott-Hubbard transition. It is not a first-order transition but a crossover where  (with decreasing $T$)  a gap first develops  at the antinodal and at  lower $T$ at the nodal point. That is with decreasing temperature, we first have a paramagnetic metal at elevated temperatures, then a pseudogap and eventually a paramagnetic insulator.

In this paper, we study the two-dimensional Hubbard model within the DF approach. Usually DF calculations are based on the local problem that arises as the solution of a converged DMFT problem. Here, we do ladder DF not only with inner self-consistency (updating the DF self-energy and Green's function) but also with outer self-consistency, which  has been studied  in parallel to this work using second order DF$^{(2)}$ \cite{Loon2018b} instead of ladder DF in our work.
We show that the two variants for the outer self-consistency  [Eqs.~(\ref{DFsc1}) and (\ref{Gupdate}) below] lead to very different results, and that for the self-consistency condition   Eq.~(\ref{Gupdate}) we can even get a metallic DF solution for large $U$'s where DMFT already yields an insulator. We also demonstrate   the exponential scaling of the correlation length in DF,
which is the cornerstone for having a paramagnetic insulator at small $U$ and low $T$.

\section{Recapitulation of the method}
\label{Sec:method}
We study the half-filled Hubbard model on a square lattice
\begin{equation}
\mathcal{H} = -t \sum_{\langle ij \rangle,\sigma} c_{i\sigma}^{\dagger}c_{j\sigma}^{\phantom{\dagger}} + U\sum_{i}n_{i\uparrow}n_{i\downarrow},
\label{eq:Hubbard}
\end{equation}
Here, $c_{i\sigma}^{\dagger}$ ($c_{i\sigma}$) creates (annihilates) an electron on site $i$ with spin $\sigma$; $\langle ij \rangle$ denotes the summation over nearest neighbors only, $U$ is the  local Coulomb repulsion and $t$ the hopping amplitude. In the following $4t\equiv 1$ sets our unit of energy.

We employ the standard ladder dual fermion method\cite{Rubtsov2009,RMPVertex}. That is, we truncate the dual fermion interaction expansion on the two-particle vertex level. Note that a full self-consistent calculation with three-particle corrections is beyond present computational resources, but higher interaction-order terms are not necessarily negligible\cite{Ribic2017b}. Hence, we calculate at least selected three-particle diagrams as an error estimate after convergence of the two-particle approach, in the same way as in Ref.~\onlinecite{Ribic2017b}.

\begin{figure}
\includegraphics[width=0.95\linewidth]{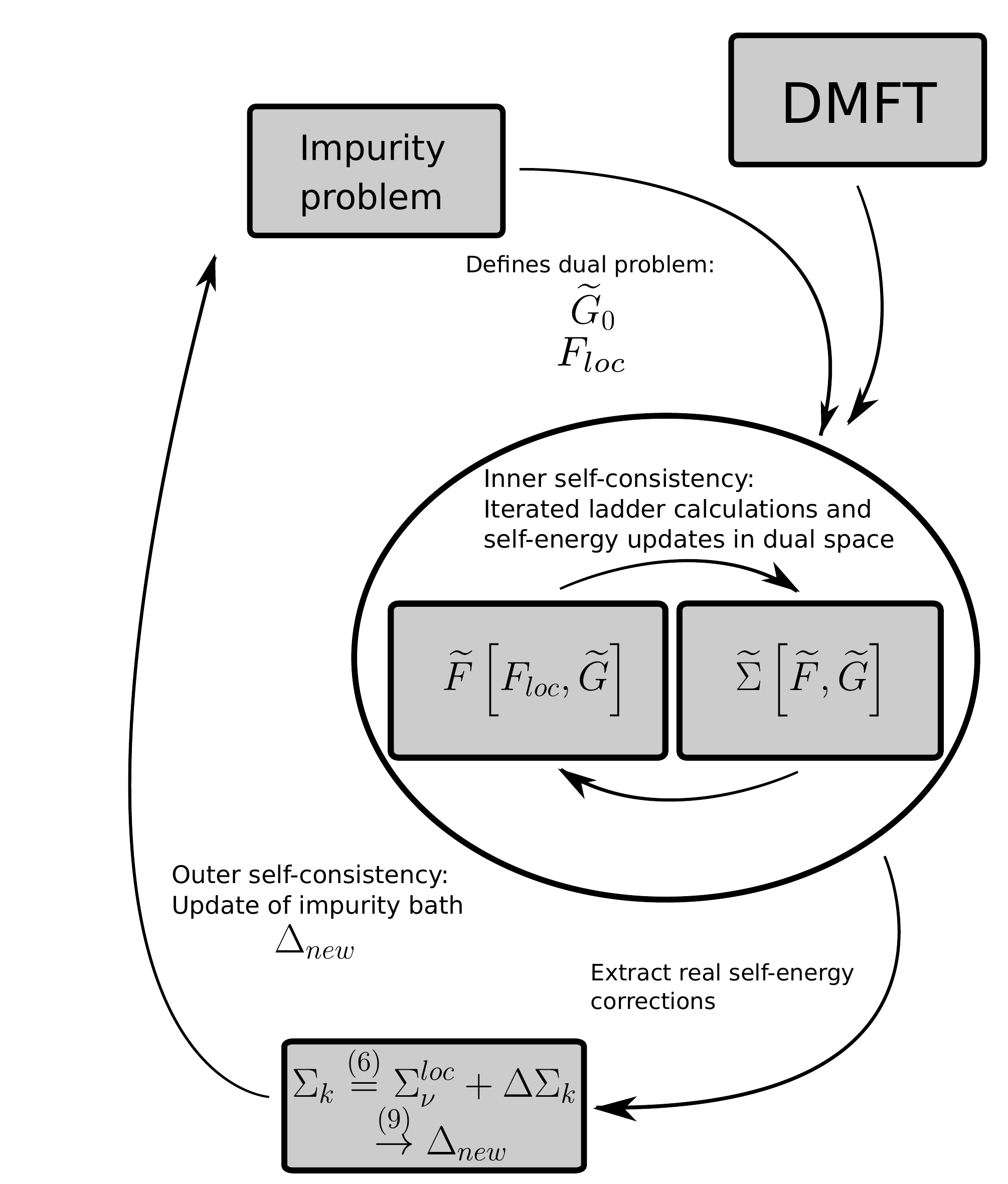} \caption{\label{Workflow}Workflow of the self-consistent DF approach. As an initial starting point we solve an impurity problem at DMFT self-consistency. This provides an initial DF interaction $F_{loc}$ and non-interacting Green's function ${\widetilde G}_0$.
In an inner self-consistency loop the DF Fermion ladder equations are solved, which yield self-energy corrections  for the real fermions. In an outer self-consistency loop also the  hybridization function is updated and a new impurity model is solved. This procedure is iterated until full self-consistently, i.e., until the impurity problem does not change any more.}
\end{figure}

 A workflow with inner and outer self-consistency  is given in Fig. \ref{Workflow}.  The calculation is started with the impurity problem of the converged DMFT solution, which yields a local full two-particle vertex $F_{loc}$ taking the place of an interaction for the dual fermions 
and a DMFT self-energy $\Sigma^{loc}$. The latter yields a $\mathbf k$-dependent physical (DMFT) Green's function 
\begin{equation}
G^{loc}_{{\mathbf k} \nu} = \dfrac{1}{i \nu - \varepsilon_{\bf{k}} - \Sigma^{loc}_{\nu} + \mu}
\label{GDMFT}
\end{equation}
and a non-interaction DF Green's function
\begin{equation}
 {\widetilde G}_{0,{\mathbf k} \nu} = G^{loc}_{{\mathbf k} \nu} - \sum_{\mathbf k} G^{loc}_{{\mathbf k} \nu},
\label{GDF}
\end{equation}
where we implicitly assumed a normalization  $\sum_{\mathbf k}=1$.

The next step in Fig. \ref{Workflow} is to calculate the interacting DF vertex  $\widetilde F$ as the series of ladder diagrams with building blocks $F_{loc}$  and interacting DF Green's function $\widetilde G$. The latter is first set to $G_{0}$ and then determined self-consistently from the DF self-energy $\widetilde \Sigma$ of the  ladder diagrams. This self-consistent solution of the ladder in dual space is called  inner self-consistency and has been routinely employed in DF calculations before \cite{RMPVertex}.

After we have achieved a converged solution in dual space, the dual self-energies are used as self-energy corrections for the real fermions. Here, we do not make use of the dual fermion mapping \cite{Rubtsov2009}
\begin{equation}
\Delta \Sigma_{k} = \dfrac{\widetilde{\Sigma}_{k}}{1 + G^{loc}_{\nu} \widetilde{\Sigma}_{k}} \; ,
\label{Eq:S1}
\end{equation}
 with  four-vector notation $k=({\mathbf k},\nu)$.
Instead, the dual self-energies are directly applied as corrections to the real fermion self-energies, 
\begin{equation}
\Delta \Sigma_{k} = \widetilde{\Sigma}_{k}.
\label{Eq:S2}
\end{equation}
This has certain advantages since the correction of Eq.~(\ref{Eq:S1}) cancels at least partially with three-particle vertex diagrams that are not considered in standard DF.
For a more detailed discussion see Ref.~\onlinecite{Katanin2013}.

In a standard DF this correction together with the local impurity self-energy
yields the physical self-energy 
\begin{equation}
\Sigma_{k}=\Sigma^{loc}_{\nu}+\Delta \Sigma_{k} \;.
\label {Eq:PSE}
\end{equation}

Here, we go beyond this standard scheme and recalculate the impurity vertex $F_{loc}$ and non-interacting DF $\widetilde G_0$ in a so-called outer self-consistency loop, see   Fig. \ref{Workflow}. Two ways to do this outer self-consistency loop have been proposed \cite{RMPVertex}. Route (i) has been previously employed and requires that 
\begin{flalign} 
&\text{(i)} \hspace{3cm} \sum_{\mathbf k} \widetilde G_{{\mathbf k},\nu}=0. \label{DFsc1}&
\end{flalign}  
 Here, we predominately follow another route (ii) instead. Route (ii) assumes that the new local impurity Green's function is given by the momentum-average over the physical Green's functions:
\begin{flalign}
&\text{(ii)} \hspace{0.5cm} G^{loc}_{new,\nu} = \sum_{\bf{k}} \dfrac{1}{i \nu - \varepsilon_{\bf{k}} - \Sigma^{loc}_{\nu} - \Delta \Sigma_k + \mu}.&
\label{Gupdate}
\end{flalign}
This impurity Green's function in turn gives a new impurity hybridization function
\begin{equation}
\Delta_{new, \nu} = \dfrac{1}{G^{loc}_{new, \nu}} - i \nu + \Sigma^{loc}_{\nu} -\mu ,
\label{Dupdate}
\end{equation}
where, $\Sigma^{loc}$ is the impurity self-energy from the previous iteration.
This $\Delta_{new}$ defines a new impurity model, which we solve using the w2dynamics\cite{w2dynamics2018,Gunacker15} continuous-time quantum Monte Carlo impurity solver in the hybridization expansion. This way we obtain   a new  $F_{loc}$ and impurity self-energy $\Sigma^{loc}$. With these we continue with Eq.~(\ref{GDMFT}) above until self-consistency. Let us remark that the repeated recalculation of the two-particle vertex $F_{loc}$ is the computational bottleneck when doing outer self-consistency. 

Please note that outside the well-investigated half-filled case, the dual corrections can change the occupation of the system as calculated from the real fermion Green's functions. Both self-consistency schemes, (i) and (ii), do not fulfill the Hartree condition $\Sigma_k\stackrel{\nu\rightarrow \infty}{\longrightarrow} \frac{U}{2}n$ where $n$ is the occupation given by the physical Green's functions of Eq.~(\ref{Gupdate}).


\section{Results}
\begin{figure}
\includegraphics[width=0.95\linewidth]{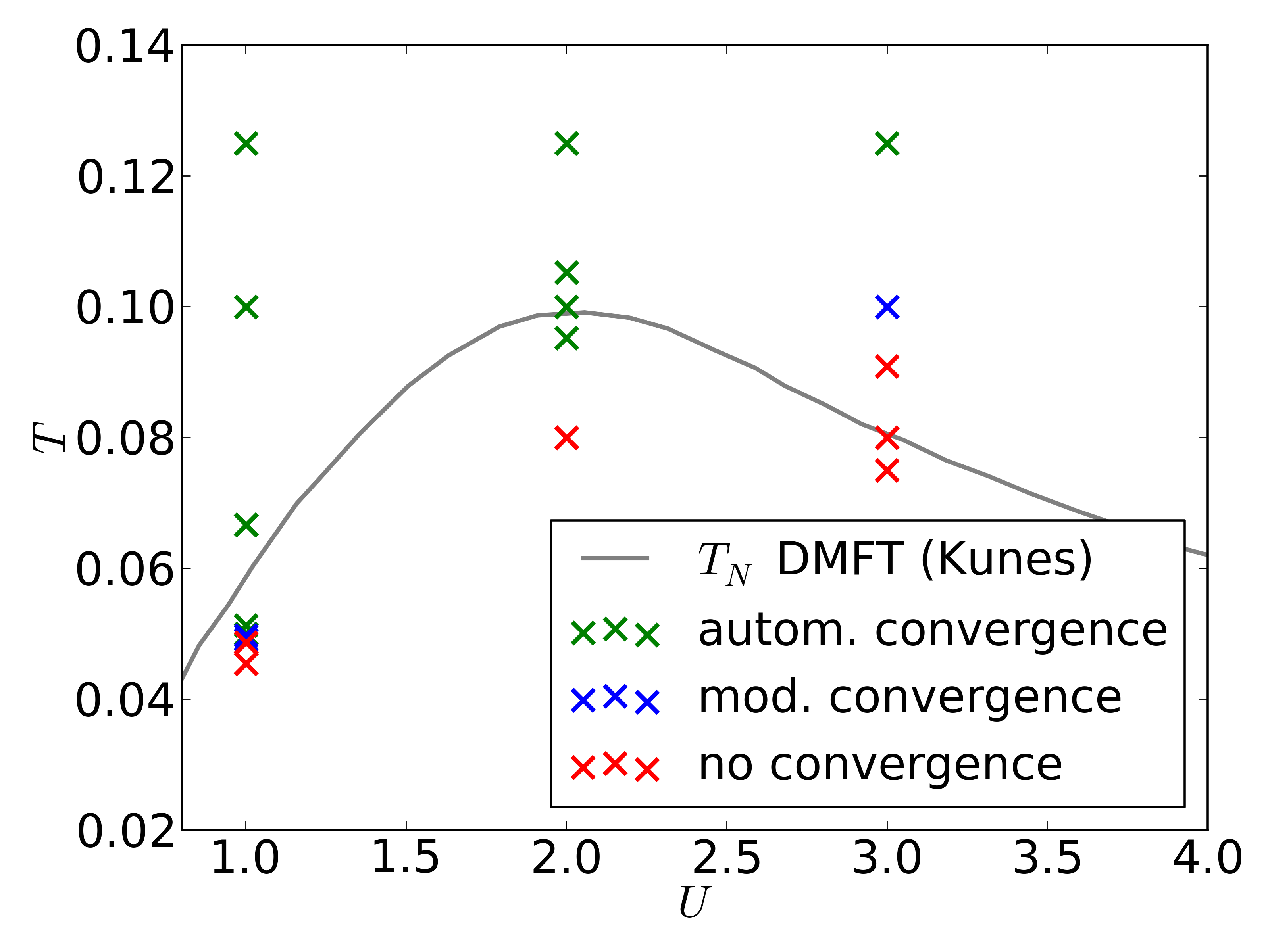}
 \caption{\label{pdiag} Data points in the phase diagram of the two-dimensional Hubbard model for which  DF calculations with outer self-consistency have been performed $(4t \equiv 1)$. The different symbols mark data points for which  convergence was achieved automatically without further modifications (green crosses), convergence was achieved after  averaging  the hybridization function of previous iterations (blue crosses) and no convergence was achieved (red crosses).
The anti-ferromagnetic ordering temperature $T_N$ in DMFT from Kune\v{s} \cite{Kunes2011} is given for comparison  (gray line).}
\end{figure}

Let us start by providing an overview in Fig.~\ref{pdiag} of which points we calculated in DF  with inner and outer self-consistency. We consider the half-filled Hubbard model on a square lattice  at relatively weak interaction ($U=1\equiv 4t$), in the DMFT strongly correlated metallic  ($U=2$) and DMFT insulating phase ($U=3$). In all cases we lower the temperature so that antiferromagnetic spin fluctuations and DF corrections become larger around the DMFT N\'eel temperature.  Similar to ladder DF with inner self-consistency only \cite{vanLoon2017}, convergence at low temperatures is not always achieved and for some parameters  convergence was only achieved after averaging the new hybridization function with the previous one using a mixing factor (under relaxation) 
, see  Fig.~\ref{pdiag}. 

\subsection{Updated impurity hybridization and self-energy}
\label{Sec:ResultsSigma}
\begin{figure}
\includegraphics[width=0.95\linewidth]{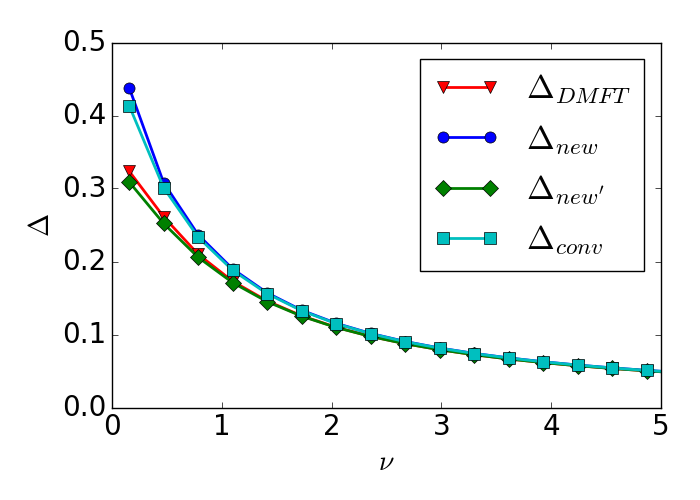} \caption{\label{JuliusDeltas} Hybridization function for $U = 1$ and $\beta = 20$ at half-filling. Depicted are the initial hybridization function at DMFT self-consistency $\Delta_{DMFT}$, the updated hybridization function after one iteration if consistency of the local Green's function is demanded $\Delta_{new}$ [scheme (ii), Eq.~(\ref{Gupdate})], the updated hybridization function after one iteration if non-locality of the dual Green's function is demanded $\Delta_{new'}$  [scheme (i) first iteration, Eq.~(\ref{DFsc1})] and hybridization function at outer self-consistency after self-consistency of the local Green's function $\Delta_{conv}$ [scheme (ii)].}
\end{figure}

Fig. \ref{JuliusDeltas} shows the change of the impurity model with outer self-consistency or more precisely the change of its hybridization function.
We find that when demanding a vanishing dual Green function [scheme (i)], the outer self-consistency  leaves the hybridization function  $\Delta_{new'}$ essentially unchanged with respect to the DMFT solution $\Delta_{DMFT}$. Shown is only the first iteration but subsequent iterations are visually not distinguishable from the first iteration on the scale of the figure.  

\begin{figure}
\includegraphics[width=0.95\linewidth]{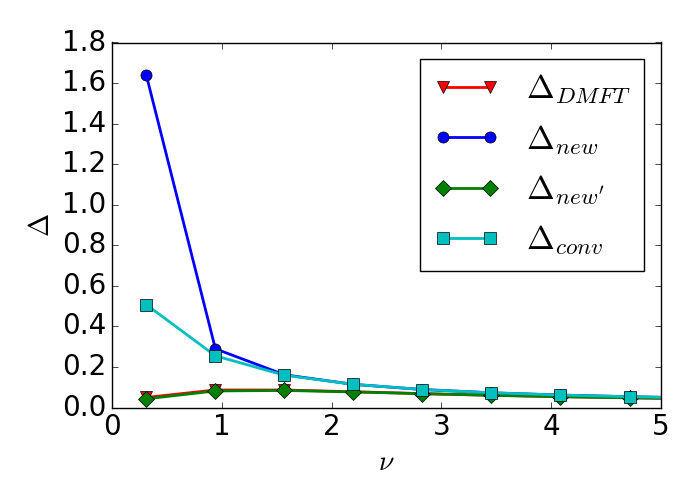} \caption{\label{Uswouldbemoreintuitive} Same as Fig.~\ref{JuliusDeltas} but for $U = 3$ and $\beta = 10$. }
\end{figure}

In contrast scheme (ii), which requires consistency between the impurity Green's function and the ${\mathbf k}$-integrated  lattice Green's function for the real fermions, shows a very different hybridization function after the first iteration ($\Delta_{new}$) but also after convergence ($\Delta_{conv}$).
In particular at  low-frequencies the hybridization function is quite considerably enhanced. Fig. \ref{Uswouldbemoreintuitive} shows that  the same tendency but to a larger quantitative extent also holds for $U=3$. Here the hybridization function is even changed from a vanishing hybridization in the low  frequency limit in DMFT and  scheme (i) to a finite one in outer-self-consistency scheme (ii).

\begin{figure}
\includegraphics[width=0.95\linewidth]{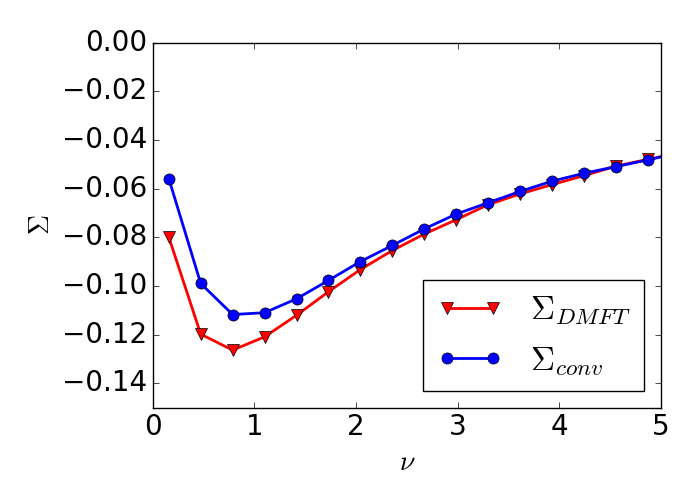} \caption{\label{JuliusSigs} Imaginary part of the  impurity self-energy for $U = 1$ and $\beta = 20$ at half-filling. Depicted are the self-energy at DMFT self-consistency $\Sigma_{DMFT}$ and the impurity self-energy at outer self-consistency $\Sigma_{conv}$ [scheme (ii)].}
\end{figure}

The enhanced hybridization corresponds to a more metallic bath or more bath states at low energies. This allows the electrons at the impurity to better evade each other, therefore reducing the imaginary part of the impurity self-energy as can be seen in Fig.\ref{JuliusSigs}.


\subsection{Local Green's functions}
 As for the impurity Green's function at low frequencies, the enhanced hybridization and reduced self-energy compete in their effect. While the reduced self-energy leads to an increase of the local Green's function at low frequencies, the stronger hybridization suppresses it. Depending on the parameter regime, the effect on the spectral function of the local system changes.

\begin{figure}
\includegraphics[width=0.95\linewidth]{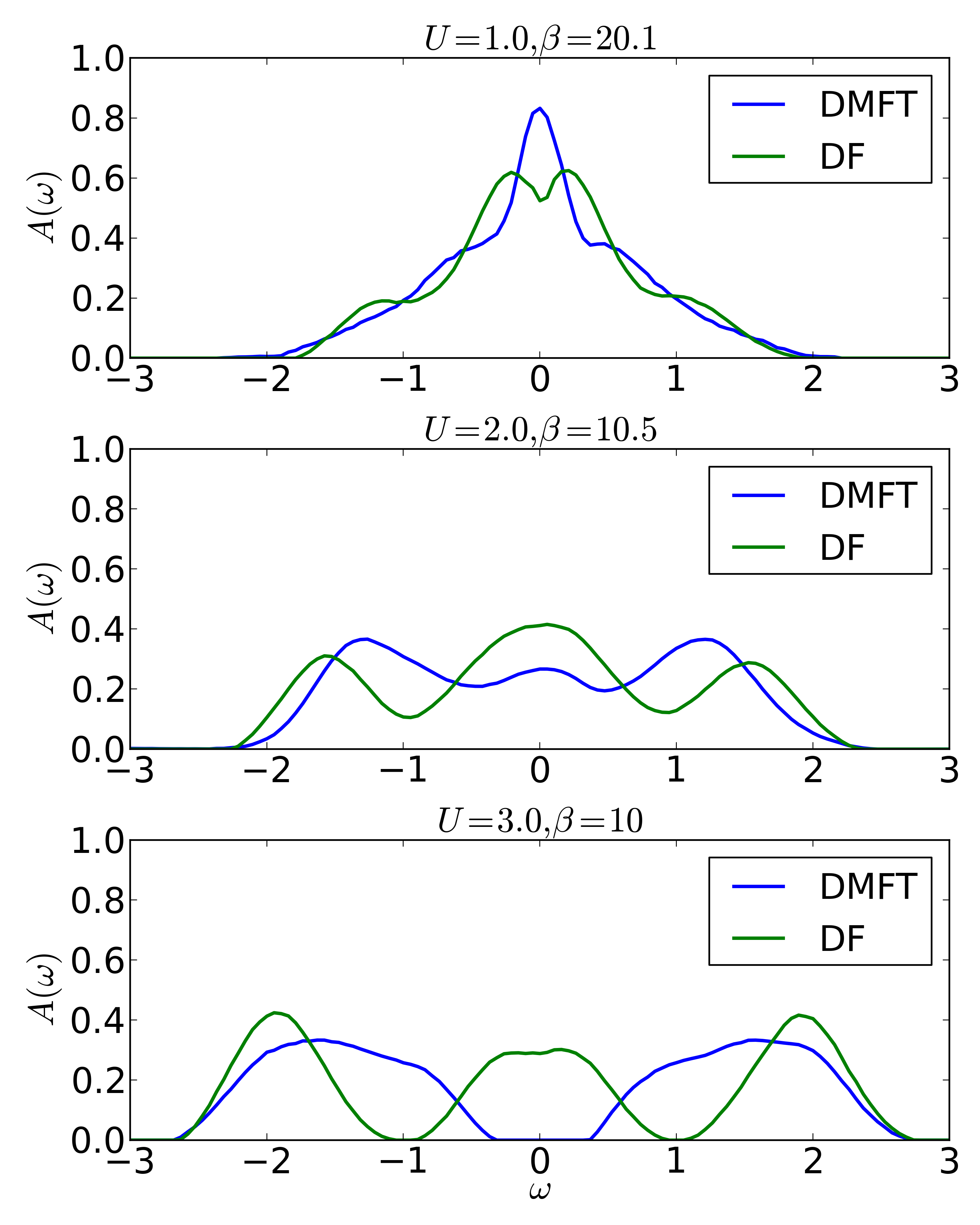} \caption{\label{spectra} Interacting density of states for selected sets of parameters at DMFT self-consistency (blue line) and  DF outer self-consistency (green line).}
\end{figure}

In Fig. \ref{spectra} we provide continuations of the local Green's functions, calculated with the maximum entropy method. The large impact of the outer self-consistency condition can be seen when comparing to the initial DMFT Green's function. Note that the impurity spectrum at self-consistency  corresponds to the lattice spectrum for scheme (ii) [see Eq.~(\ref{Gupdate})]. For $U=1$, the non-local correlations lead to a more insulting behavior of the DF solution. On the other hand, for $U=3$,
the DF solution is ---at least at $\beta=10$--- metallic with a three-peak spectrum, while the DMFT solution is already in the Mott insulating phase.  

This is counter-intuitive since one expects non-local spin fluctuations to result in a more insulting solution. It is a consequence of the larger (more metallic) hybridization function in  scheme (ii), see Fig.~\ref{Uswouldbemoreintuitive}, which pushes the  $U=3$ impurity model  into the metallic phase. The origin of the larger hybridization in turn is that, because of the $\mathbf k$-dependent DF self-energy, the spectrum at occupied and unoccupied  $\mathbf k$-points is pushed further away from each other, as we will see in the next section. In an impurity model we can only describe this larger spectral width at fixed $U$ if we have a larger hybridization function. 

\subsection{Comparison to DCA}
\begin{figure}
\includegraphics[width=0.95\linewidth]{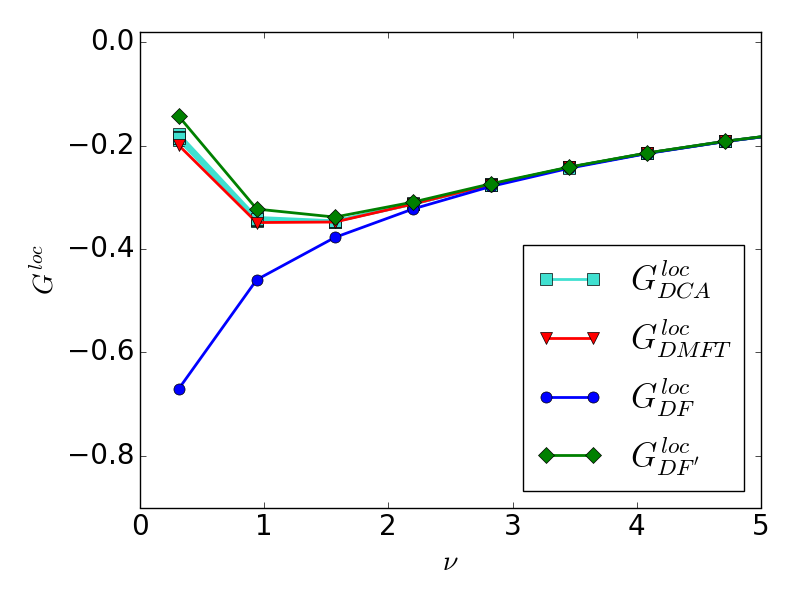} \caption{\label{QuenyaGlocsDCA} Local Green's functions for $U=3$ and $\beta=10$ within $DF$ [scheme (i)], $DF'$  [scheme (ii)] and DCA with cluster sizes 4, 8, 16, 18 and 32 (indiscernible on the scale of the figure).} 
\end{figure}

\begin{figure}
\includegraphics[width=0.95\linewidth]{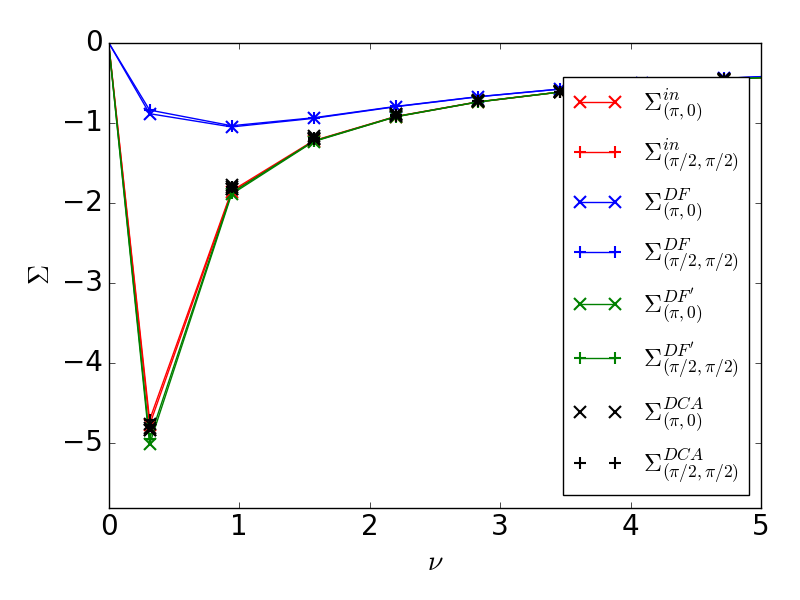} \caption{\label{QuenyaSigsDCA} Self-energies for nodal and antinodal $\bf{k}$-points for $U=3$ and $\beta=10$ within $DF$ [scheme (i)], $DF'$ [scheme (ii)] and DCA with cluster sizes 4, 8, 16 and 32 (which lie essentially on top of each other).} 
\end{figure}

Since it is doubtful whether these results of outer-self-consistency scheme (ii) are describing the correct physics, we have compared the results with DCA  Green's functions and self-energies on the Matsubara axis, see Fig.  \ref{QuenyaGlocsDCA} and \ref{QuenyaSigsDCA}, respectively. The DCA yields an insulating spectrum as is indicated by a vanishing $G^{loc}$ for frequency $\nu \rightarrow 0$ in  Fig.~\ref{QuenyaGlocsDCA} and a divergent  $\Sigma$ for $\nu \rightarrow 0$ in  Fig.~\ref{QuenyaSigsDCA}. This is qualitatively and even quantitatively the same behavior as in DMFT as well as in  DF without outer self-consistency (and also the outer self-consistency scheme (i) hardly changes the hybridization and hence this result).  Clearly DF with outer self-consistency scheme (ii) is off. 

This puts severe doubts on the self-consistency scheme (ii). A possibility is that third-order diagrams cure this effect as they provide an extra term proportional to $1/(i\nu)$ which, in principle, could provide a more insulating solution again. But without further modifications the outer-self-consistency scheme (ii), requiring that the impurity Green's function equals the physical Green's function, does not properly work at large $U$.
Let us also note, that employing the translation from DF self-energies to real ones using Eq.~\eqref{Eq:S1} instead of Eq.~\eqref{Eq:S2} hardly affects the results since either the self-energy or the Green's function in the denominator of Eq.~\eqref{Eq:S1}
is small.

\subsection{DF and physical self-energies} 
\begin{figure}
\includegraphics[width=0.95\linewidth]{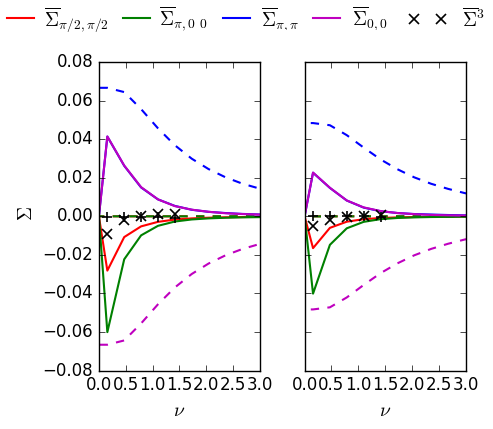} 
\caption{\label{JuliusAlexanderSigmas} Dual self-energies for different $\bf{k}$-points for $U = 1$ and $\beta = 20$ at inner self-consistency only (left; DMFT impurity problem) and inner and outer self-consistency (right). Imaginary parts are depicted as full lines and real ones as dashed lines. Additionally, momentum-independent three particle corrections $\overline{\Sigma}^3$ are given, with ``$+$''-symbols for the real and ``$\times$''-symbols for the imaginary part.}
\end{figure}

Next, let us discuss the resulting self-energies, real as well as dual ones. The imaginary part of the impurity self-energies $\Sigma^{loc}$  was already shown in  Fig.~\ref{JuliusSigs} and is found to be consistently reduced by the employed outer self-consistency scheme. $\Sigma^{loc}$ already represents the major contribution to the self-energy for the real fermions, with the dual fermion self-energy shown in Fig.~\ref{JuliusAlexanderSigmas} yielding only quantitatively smaller corrections. Note that the physical self-energies are just given by the sum of 
 Fig.~\ref{JuliusSigs} and  Fig.~\ref{JuliusAlexanderSigmas} according to Eq.~(\ref{Eq:PSE}).

 We also find that the change of $\Sigma^{loc}$ throughout the iterations is quantitatively larger than the corrections due to the dual fermion calculations themselves. The dual self-energies for {\bf{k}}-points on the Fermi-edge lead to enhanced scattering and reduce the contribution of these states to the low-frequency spectral function. For points sufficiently far from the Fermi-edge the sign of the imaginary part of the dual self-energy changes from negative to positive, corresponding to a reduction of the scattering rate. Also, the dual corrections pick up a real part, shifting the states such as ${\mathbf{k}}=(0,0)$ and ${\mathbf{k}}=(\pi,\pi)$  further away from the Fermi energy. Consequently, the overall width of the spectrum is enhanced.

\begin{figure}
\end{figure}

\subsection{Susceptibilities and correlation lengths} 
\begin{figure*}
\begin{minipage}{0.32\linewidth}
\includegraphics[width=0.95\linewidth]{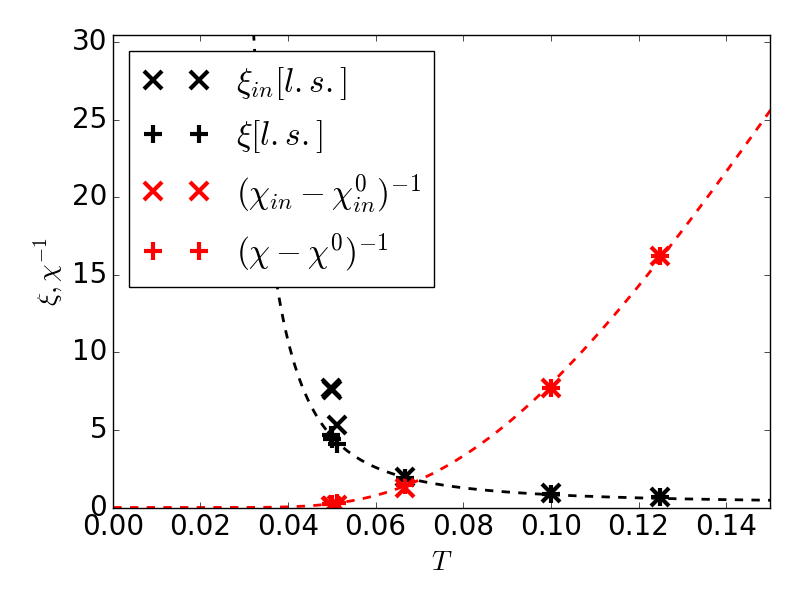}
\includegraphics[width=0.95\linewidth]{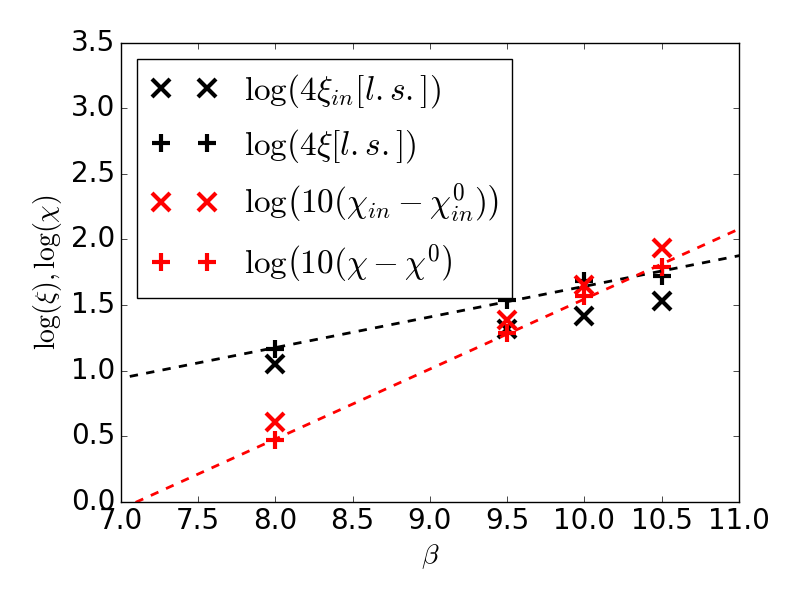}
\end{minipage}
\hfill
\begin{minipage}{0.32\linewidth}
\includegraphics[width=0.95\linewidth]{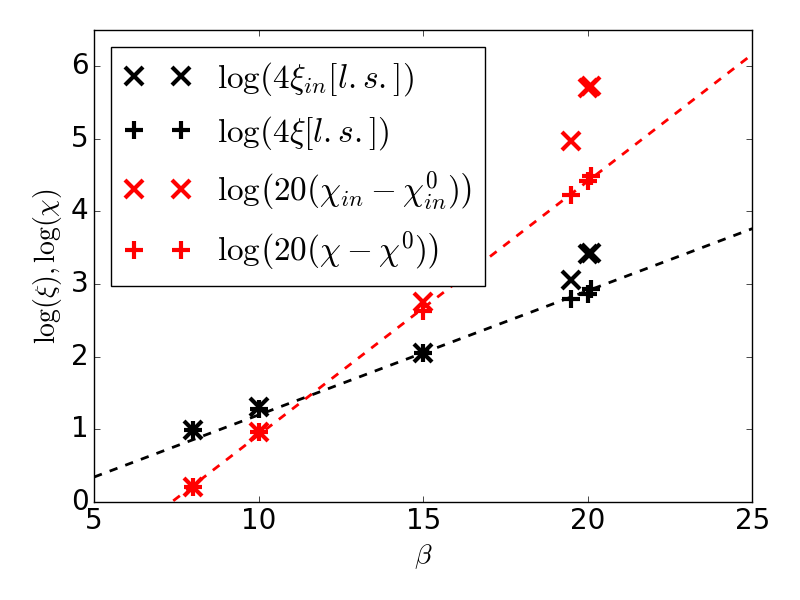}
\includegraphics[width=0.95\linewidth]{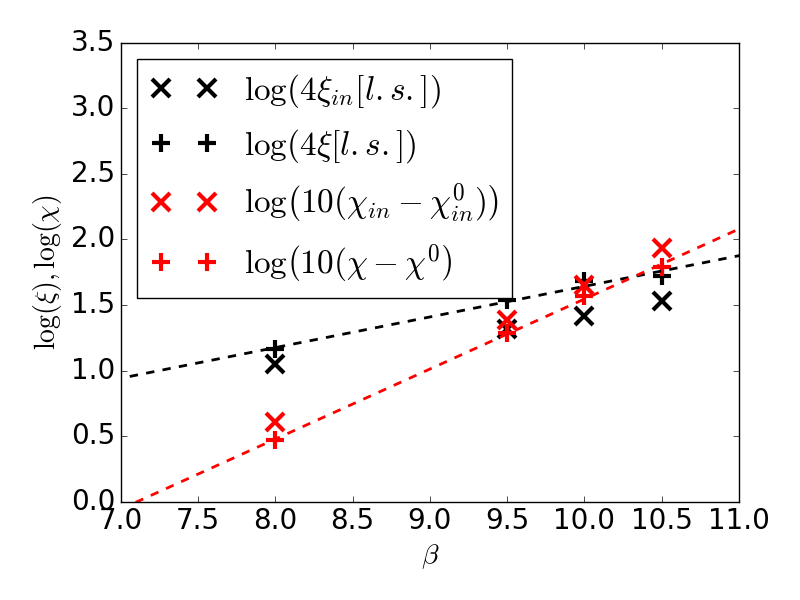}
\end{minipage}
\hfill
\begin{minipage}{0.3\linewidth}
 \caption{\label{TU1} Left: (Inverse) vertex contribution to the antiferromagnetic susceptibility $(\chi-\chi^0)$ and correlation length $\xi$ extracted from it a function of temperature for $U=1$ (upper panel) and $U=2$ (lower panel) at half-filling. The light symbols denote the results obtained from the initial DMFT impurity model at inner dual self-consistency only ($in$); the dark symbols those after outer self-consistency is achieved. Right: Same as left but on a logarithmic scale and as a function of $\beta=1/T$. The correlation length is measured in lattice spacings (l.s.) and the remaining susceptibilities are in units of $\mu_B^2$. 
 }
\end{minipage}
\end{figure*}

Let us finally turn to the DF susceptibilities and correlation lengths and their change with outer self-consistency shown in Fig.~\ref{TU1}. We focus here on the vertex contribution to the susceptibility which becomes dominant for $T \lesssim 0.07 \, (U=1)$ and $T \lesssim 0.1 \, (U=2)$. Only plotting the vertex contribution has the advantage that the exponential behavior can be seen up to higher temperatures. The additional bare bubble susceptibility has a weaker temperature dependence and hence becomes dominant/obfuscates the exponential behavior of the vertex contribution. 
At low interaction values, the outer self-consistency scheme (ii) is found to reduce the antiferromagnetic susceptibility for the antiferromagnetic {\bf{q}}-point $(\pi , \pi)$. Also, correlations lengths extracted from the width of the (vertex part of the) magnetic susceptibilities on a {\bf{q}}-mesh are shorter when based on the impurity with outer-self-consistent local Green's functions compared to the calculation based on the DMFT impurity. A similar trend is found for the inner self-consistency: it suppresses the antiferromagnetic susceptibility which otherwise would diverge at the DMFT $T_N$  (because  the Green's function becomes more damped with inner self-consistency). In the case of outer self-consistency the physical reason is the enhanced hybridization strength. Because it also gives---unphysically---a more metallic solution at large $U$, it is however doubtful whether this is an artifact of the self-consistency scheme (ii).

Plotting vertex contributions to the susceptibility and correlation lengths on a logarithmic scale as a function of the inverse temperature $\beta$, as it is done in Fig. \ref{TU1} (right) shows a clear linear trend which translates into the exponential scaling $\chi \sim e^{a/T}$ and  $\xi \sim e^{b/T}$, both with and without outer self-consistency. This is the first time that exponentially large correlation lengths have been demonstrated also for DF.

\section{Conclusion}
\label{Sec:conclusion}
We have shown that enforcing outer self-consistency requiring the physical and the impurity Green's function to be equivalent [scheme (ii)] changes the results of dual fermion calculations not only quantitatively, but in some cases also has a qualitative effect and causes new phenomena, such as the appearance of a three-peak metallic spectrum at $U=3$, above the $U_c$ of the DMFT Mott transition. The reason for this is the larger hybridization function of the self-consistent impurity problem which needs to accommodate a larger spread of the spectral function because the $\mathbf k$-dependent DF self-energy pushes states further away from the Fermi energy. The self-consistency scheme (i), which requires a purely non-local  interacting dual Green's function, on the the hand, hardly changes the hybridization function so that corrections due to the outer self-consistency scheme (i) are small when using the ladder DF approach.
 This  poses pressing questions about the best outer self-consistency scheme and its  underlying impurity problem. Our results indicates that scheme (ii) might be inferior but for a conclusive answer further benchmarks against independent methods, such as cluster Monte-Carlo simulations are called for.

Another important result of our paper is that the DF approach also yields an exponential scaling of the correlation length at low temperatures for the two dimensional Hubbard model. We were able to demonstrate this by focusing on the vertex contribution to the susceptibility, which becomes dominant for low enough temperatures. This reaffirms the scenario that the paramagnetic  phase is always insulating at low enough temperatures and for a lattice with perfect nesting \cite{Vilk1996,Vilk1997,Schaefer2015-2}, essentially because the spectrum looks (almost) like that of the antiferromagnetic ground state if the correlation lengths are that large. In contrast, for second-order DF$^{(2)}$ with outer self-consistency a finite $U_c$ was found.\cite{Loon2018b} With the strongly increasing correlation length in our ladder DF calculations, one might also expect that the metallic DF solution at $U=3$ eventually becomes insulating due to strong antiferromagnetic spin fluctuations, but only at much lower temperatures.

\section*{Acknowledgments}

We thank  E.G.C.P van Loon, G. Rohringer, A. Rubtsov and A. Toschi for helpful discussions.
We thank J. Kaufmann for providing a MaxEnt reference code.
We are particularly grateful to E. Gull and A. Chen for providing DCA results for comparison.
The local impurity problems were solved in w2dynamics\citep{w2dynamics2018,Gunacker15} (CT-HYB); 
the plots were made using the matplotlib \cite{Matplotlib} plotting library for python. Financial support is acknowledged from the European Research Council under the European Union's Seventh Framework Program (FP/2007-2013)/ERC through
grant agreement n.\ 306447 (TR, KH). PG has been supported by the
Vienna Scientific Cluster (VSC) Research Center funded by the Austrian
Federal Ministry of Science, Research and Economy (bmwfw).
The computational results presented have been achieved using the VSC and computational resources provided by XSEDE grant no. TG-DMR130036.

{\em Note added.} During the completion of this work, we learned about another DF outer-self-consistency study using DF$^{(2)}$ and focusing on self-consistency scheme (i).\cite{Loon2018b}

\appendix



\end{document}